\documentclass[twocolumn,aps,showpacs]{revtex4}

\usepackage{graphicx}
\usepackage{dcolumn}
\usepackage{bm}
\usepackage{float}

\begin{document}


\title{Flux surface shaping effects on tokamak edge turbulence and flows}

\author{Alexander Kendl}%
\affiliation{%
Association EURATOM-\"OAW, Institut f\"ur Theoretische Physik, Universit\"at
Innsbruck, A-6020 Innsbruck, Austria
}%
\author{Bruce D. Scott}%
\affiliation{%
Max-Planck-Institut f\"ur Plasmaphysik, EURATOM Association, D-85748
Garching, Germany
}%

\date{\today}

\begin{abstract}
Shaping of magnetic flux surfaces is found to have a strong impact
on turbulence and transport in tokamak edge plasmas.
A series of axisymmetric equilibria with varying elongation and
triangularity, and a divertor configuration are implemented into a
computational gyrofluid turbulence model.  
The mechanisms of shaping effects on turbulence and flows are
identified.
Transport is mainly reduced by local magnetic shearing and an enhancement of
zonal shear flows induced by elongation and X-point shaping.
\end{abstract}

\maketitle

\section{Introduction}

It can be regarded as one of the most intriguing results of half a
century in
physics of hot magnetized plasmas, that in macroscopically stable equilibria
the confinement is nevertheless determined by micro-scale
instabilities and fluctuations, with
transport of energy and particles across
magnetic flux surfaces being dominated by ubiquitous turbulence on small
(gyro radius) scales and low (drift) frequencies
\cite{Tang78,Hugill83,Liewer85,Wootton90,Dimits00,Scott03PPCF,Garbet04}.
This turbulence is regulated by the formation of
mesoscopic zonal structures out of the turbulent flows
\cite{Waltz94,Zlin98,Terry00,Hahm00,Hahm02,ZFReview}.  

Criteria for design of new magnetized plasma experiments for fusion research
like ITER are primarily based on empirical scaling laws for the energy
confinement time \cite{ITER03}.
First-principle based transport models still require reference to
experimental scalings, and are validated mainly only for core plasmas
excepting the pedestal region \cite{Dimits00}.  
An improved understanding of the underlying turbulent dynamics and of its
relation to design parameters, like the shape of the confining magnetic
field, will facilitate the development of advanced magnetic confinement
experiments.

The plasma shape of a tokamak enters into confinement and transport
modelling through parameters specifying a vertical elongation $\kappa
\geq 1$
and an outboard side triangularity $\delta \geq 0$ that describe the
deviation
from a simple circular torus. The design criteria of the ITER plasma shape
with $\kappa = 1.7$ and $\delta=0.33$ (at $95 \%$ flux
surface) are based on a conservative regime that is well established
in present experiments \cite{ITER03}.
More extreme variations of flux surface shaping were investigated with
the experiment TCV (Tokamak \`a Configuration Variable) which is able
to achieve configurations up to $\kappa=3$ and $\delta=\pm 0.5$
\cite{TCV97}.
Experimental evidence suggests that large elongation is always
beneficial for local and global energy confinement, whereas the role
of triangularity depends also on the value of the plasma pressure
gradient.

\begin{figure}[H]
\includegraphics[width=8.5cm]{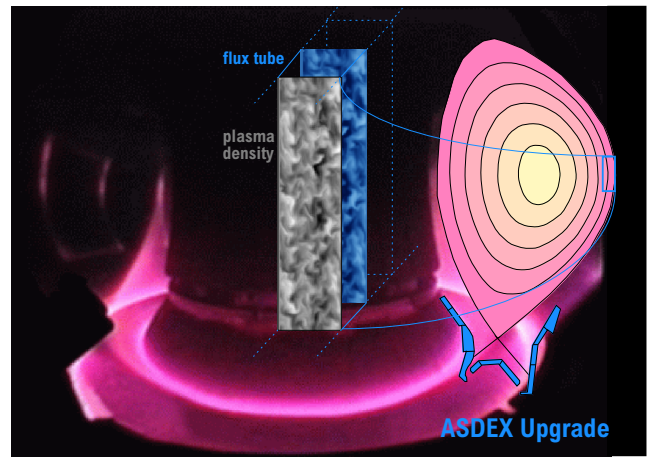}
\caption{\label{f:aug} Poloidal cross section of a tokamak plasma
(ASDEX Upgrade) showing the elongation and triangularity of closed
flux surfaces and a lower X-point on the separatrix.
A section of the flux tube for drift wave turbulence computations
winding around the torus along field lines is here locally intersecting the
outer midplane of the plasma edge.}
\end{figure}

An important task for computational plasma physics is to determine the
effects of flux surface shaping on turbulence and transport by
first-principle direct numerical simulation.
Shaping effects are complicated even for apparently simple circular
equilibria \cite{Scott97a}, and efforts to perform modelling for edge
plasmas most often incorporate technically complicated prescriptions in
an inflexible way \cite{Xu}.  An advantageous way to explore shaping is
through the use of a covariant metric in a model with Cartesian topology
\cite{ScottVar98,Scott01}, but even then it was found that the use of
experimental geometry, while most realistic for modelling purposes, may not be
helpful to theoretical understanding, and hence standard ``kappa-delta''
models were used \cite{Scott2000}.  In attempt to understand the effects
of stellarator geometry, a set of analytical forms were used for the
metric, varying each in turn to provide a small set of steps from
tokamak to stellarator geometry \cite{Kendl00}.  Forms like this were
also used to characterise the X-point effects, again making
simple control tests tractable \cite{Kendl03}.

In the following, we report on the results of a more systematic series
of ``kappa-delta'' tokamak configurations
with varying elongation and triangularity, and an actual ASDEX Upgrade
\cite{AUG} divertor configuration with the equilibrium code HELENA
\cite{HELENA}. These are implemented into the gyrofluid code GEM
(\cite{GEM}, which is updated from Ref.\ \cite{Scott2000})
for computations of edge turbulence and flows.
Specific effects of flux surface shaping on drift-Alfv\'en (DALF) and ion
temperature gradient (ITG) turbulence in the tokamak edge plasma are
analysed
and compared, and the mechanisms of the coupling between geometry and
turbulence are identified.  The emphasis is on physical mechanism, which
is complimentary to and should also be helpful to the understanding of
the more experimentally oriented modelling efforts.

It is found that turbulent transport fluxes are reduced by increasing flux
surface elongation $\kappa$. The influence of triangularity $\delta$ is much
weaker and generally also depends on elongation and on a pressure shift.
The geodesic acoustic oscillation modes (GAM) and the overall turbulence
frequency spectra depend on flux surface geometry.
An enhancement of zonal flow amplitude in the spectra by elongation and
X-point shaping is found to be weak in the DALF case and more pronounced in
the ITG case. The reduction of transport observed in our simulations in
shaped tokamak geometry is mainly a result of local and global magnetic
shear.

We begin in Section II with a brief review and discussion of the role of
sheared flows in tokamak turbulence in context of the toroidal flux surface
geometry.
In section III the gyrofluid electromagnetic model GEM for edge turbulence
computations is presented, and the flux tube representation of tokamak
geometry within a GEM turbulence code is discussed in Section IV. Transport
scalings by elongation and triangularity, that are obtained with this model,
are presented in Section V. Resulting frequency and wave number spectra in
tokamak divertor geometry are discussed in Section VI. Conclusions and
implications of the results are presented in Section VII.

\section{The role of sheared flows in tokamak turbulence}

One of the main requirements for successful performance of
fusion experiments is the prospect of operation in a high
confinement H-mode \cite{Wagner82,Gohil94,Suttrop97}.
The formation of a characteristical transport barrier in
magnetically confined toroidal H-mode plasmas is closely related to
the presence of a radially varying electric field near the last closed
flux surface \cite{Groebner90,Gohil94}.  A radial electric field, $E_r$,
gives rise to a perpendicular E-cross-B (ExB) flow through the relation
${\bf v}_{E} = (c/B^2) {\bf E} \times {\bf B}$.  If $E_r$ varies
with $r$, then this flow is sheared, and this flow shear is what is
invoked to explain the transport barrier
\cite{Groebner90,Gohil94,Burrell97}.
The flow shear (vorticity)
is expected to suppress turbulent transport either by
shearing the eddies apart \cite{Biglari90} or by simply tilting them so
that their Reynolds-stress interaction with the vorticity layer tends
towards energy transfer out of the eddies and into the flow
\cite{DiamondKim91}.  Neoclassical equilibrium effects can also produce
these shear layers \cite{McCarthy93,Heikkinen00}.  As opposed to these
mean ExB flows, the fluctuating zonal flows in direct interaction with
the turbulence have also been observed in tokamak edge regions
\cite{Shats03}.

The transition between states of low (L) plasma confinement to the
high (H) confinement regime featuring such an edge transport barrier
is up to now not satisfactorily described by any of the existing
first-principle theories \cite{Connor00}.  
The notion that generation of zonal flows by direct drive of the
turbulence can act to self-regulate by suppressing the driving
turbulent vortices has lead to the development of predator-prey type
bifurcation models that are able to describe specific characteristics
of the L-H transition (see e.g. \cite{Diamond94}).
Early computations based on the resistive-g \cite{Carreras} and
collisional drift wave \cite{IAEA94} models found self-generated sheared
flows which could be argued as an L-H transition trigger. However,
more recent models in more comprehensive treatment of the toroidal
geometry and also the full range of scales of motion do
not find this \cite{Scott2000}.
The phenomenology of these self consistent zonal flows receives wide
interest \cite{Hahm00,Hahm02,ZFReview}, but they serve as a moderator of the
turbulence rather than a suppression mechanism, as the energetic
interactions with
not only the pressure but also magnetic parts of the equilibrium tend to
place each part of the overall dynamical system, which naturally
includes the flows themselves \cite{Scott05}.

After the initial discovery of the H-mode in the divertor experiment ASDEX
\cite{Wagner82} it has been found in numerous large tokamaks around
the world that the presence of a divertor (originally designed to improve
the impurity exhaust) significantly enhances the prospect of reaching this
state \cite{Carlstrom96}.
However, up to now both the transition models as well as the numerical
simulations have relied mainly on simplified representations of the plasma
geometry. In our present work we investigate the effects of realistic flux
surface shaping on tokamak edge turbulence using a gyrofluid model recently
corrected for energy conservation properties \cite{Scott03PPCF,GEM}, thereby
updating previous gyrofluid results \cite{Scott2000}.
This provides a
counterweight to existing efforts based on collisional fluid equations
and associated numerical schemes
which cannot treat the ion gyroradius scales and therefore overestimate
linear phenomena based in the larger scales \cite{Xu,Kleva}.
The ion gyroradius necessarily enters because the dynamics is sensitive
to phenomena in the range of the spectrum with perpendicular wavenumbers
only slightly less than the inverse of the drift scale ($\rho_s$, also
called ``ion sound gyroradius''), and with comparable ion and electron
temperatures $\rho_s$ is commensurate with the ion gyroradius $\rho_i$
\cite{Scott03PPCF}.
A schematic view of a shaped plasma cross section including a flux-tube
section from turbulence computations is shown in Fig.~\ref{f:aug}.  

Transport in a fusion edge plasma is dominated by turbulent low-frequency
drift wave motion that causes a fluid-like convection through ExB
vortices in a plane perpendicular to the magnetic field direction,
acting in concert with parallel coupling dynamics tending towards
dissipation \cite{HasMim78,Horton81,WakHas84}.
In general this is both nonlinear and electromagnetic
\cite{Scott03PPCF,Garbet04}.
Toroidal compressibility of zonal (flux surface averaged) ExB flows
opens channels in the nonlinear dynamics of the vortex-flow
interaction that strongly affect the flow energetics
\cite{Hallatschek,Scott03}.  
The origin of the compression is the geodesic curvature of the
magnetic field lines. The flow is thereby coupled
with poloidally asymmetric pressure sidebands which are
consumed by the turbulence and global Alfv\'en or parallel flow
dynamics.  This energetics was given in Ref.\ \cite{Scott03} and
the various coupling effects throughout the flow/equilibrium system were
analysed in Refs.\ \cite{Scott05,Naulin05},
and the sensitivity of the turbulence to the geodesic energy transfer
effect was explored using an additional scaling parameter for the
strength of its coupling \cite{Kendl05}.

This geodesic transfer mechanism represents
a restoring loss channel for the zonal flows, ultimately placing them in
statistical equilibrium with the turbulence, with the Reynolds stress
(spin-up) mechanism continuing to operate.  Since the latter acts on all
frequencies, the geodesic acoustic oscillation (GAM) \cite{Winsor}
itself need not be present for the geodesic transfer mechanism to
operate. The net result of this is that turbulent transport in 3D  
toroidal edge computations is found to be reduced but not completely
suppressed by the self-generated zonal flows.
This geodesic transfer effect was first studied for the case of a
simple circular toroidal magnetic field \cite{Scott03,Scott05,Naulin05},
In the following we give both numerical evidence and a physical picture
of how
plasma shaping and the presence of a divertor via its modification of plasma
geometry can influence the outcome of the gyrofluid turbulence/flow
interaction.

 \section{The gyrofluid electromagnetic model GEM}

The dynamical character of cross-field ExB drift wave turbulence in the edge
region of a tokamak plasma is governed by electromagnetic and dissipative
effects in the parallel response.  
The most basic drift-Alfv\'en (DALF) model to capture the drift wave
dynamics
includes nonlinear evolution equations of three fluctuating fields: the
electrostatic potential $\tilde \phi$, electromagnetic potential $\tilde
A_{||}$, and density $\tilde n$. The tokamak edge usually features a more or
less pronounced density pedestal, and the dominant contribution to the free
energy drive to the turbulence by the inhomogeneous pressure background is
thus due to the density gradient. On the other hand, a steep enough ion
temperature gradient (ITG) in the edge does not only change the turbulent
transport quantitatively, but adds new interchange physics into the
dynamics. In addition, more field quantities have to be treated:
parallel and perpendicular temperatures $\tilde T_\parallel$ and $\tilde
T_\perp$ and the associated parallel heat fluxes, for a total of six
moment variables for each species. Finite Larmor radius effects
introduced by
warm ions require a gyrofluid description of the turbulence equations.  
Both the resistive DALF and the ITG models can be covered by using the
six-moment electromagnetic gyrofluid model GEM \cite{GEM}, but we will refer
for the DALF model to its more economical two-moment version
\cite{Scott03PPCF}.

The gyrofluid model is based upon a moment approximation of the
underlying gyrokinetic equation.  The first complete six-moment
formulation was given for slab geometry \cite{Dorland93}, and later
extended to incorporate toroidal effects \cite{Beer96a} using a
ballooning-based form of flux surface geometry \cite{Beer95}.
Electromagnetic induction and electron collisionality were then included
to form a more general gyrofluid for edge turbulence \cite{Scott2000},
with the geometry correspondingly replaced by the version from the edge
turbulence work, which does not make ballooning assumptions and in
particular represents slab and toroidal eigenmode types equally well and
does
not require radial periodicity \cite{Scott01}.  Energy conservation
considerations were solidified first for the two-moment version
\cite{Scott03PPCF}, and most recently for the six-moment version
\cite{GEM}.  This latter reference currently defines the GEM model,
energetics, and also the numerical techniques.  The present paper uses
the six-moment equation set for each species (Eqs.\ 99-104),
induction and polarisation equations (Eqs.\ 89,92), the zero gyroradius
limit for electrons, the local model with separate linear drive terms
and Dirichlet boundary conditions (Sec.\ IV A), Pad\'e forms for
gyroaveraging operators (Sec.\ IV B), and the dissipation model and
numerical scheme (Secs.\ IV C-E), all with equation and Section numbers
from Ref.\ \cite{GEM}.  This is the standard local form for the GEM code.

The flux surface geometry of a tokamak enters into the gyrofluid equations
via the curvilinear generalisation of differentiation operators and via
inhomogeneity of the magnetic field strength $B$. The different scales of
equilibrium and fluctuations parallel and perpendicular to the magnetic
field
motivate the use of field aligned flux coordinates.
The differential operators in the field aligned frame are the parallel
gradient  
\begin{equation}
\nabla_\parallel = (1/B) ({\bf B} + \tilde{\bf B}_\perp) \cdot \nabla,
\end{equation}
with magnetic field disturbances $\tilde{\bf B}_\perp = (-1/B) {\bf B}
\times \nabla \tilde A_\parallel$ as additional nonlinearities,
the perpendicular Laplacian
\begin{equation}
\nabla_\perp^2=\nabla\cdot[(-1/B^2){\bf B}\times({\bf B}\times\nabla)],
\end{equation}
and the curvature operator
\begin{equation}
{\cal K}=\nabla\cdot[(c/B^2){\bf B}\times\nabla)].
\end{equation}

The dynamical character of the system is further determined by a set of
parameters characterising the relative role of dissipative, inertial and
electromagnetic effects in addition to the driving by gradients of
density and
temperature.  In particular, we specify
collisionality $C = 0.51 \hat\epsilon (\nu_e c_s/L_\perp) (m_e/M_i)$,
magnetic induction $\hat \beta = \hat\epsilon (4 \pi p_e / B^2)$,
electron inertia $\hat \mu = \hat\epsilon (m_e/M_i) $
and ion inertia $\hat\epsilon = (qR / L_{\perp})^2$,
where $L_{\perp}$ is the background gradient scale length, and
$c_s=\sqrt{T_e/M_i}$ is the sound speed. We use a standard set of edge
parameters given as $C=5$, $\hat \beta = 1, \hat \mu = 5$ and $\hat
\epsilon=18350$.  

The perpendicular scale length for the DALF model is set as $L_{\perp} =
L_n$
to the density gradient length, and for the ITG model as $L_{\perp} =
L_{Ti} =
0.5 L_n = 0.5 L_{Te}$ to the ion temperature gradient length, so that
$\eta_i
= L_n / L_{Ti} = 2$. The ITG model has otherwise identical
parameters. Accurate experimental measurement of ion temperature profiles in
the tokamak pedestal region are only recently being advanced and have
previously been plagued by large uncertainties. The present experimental
knowledge of edge parameters can thus be seen commensurate with both the
DALF and ITG model cases used here.

The computational domain is set to $64 \times 256$ nodes in units of the
drift scale  $\rho_s=(c/eB)\sqrt{T_eM_i}$ for $(x,y)$ and 16 nodes in
one field line connection length ($2\pi qR$) in $-\pi<z<\pi$.

The dimensions are chosen to appropriately account for statistical
isotropy in small scales in both perpendicular directions with
satisfactory spectral overlap, and for an extended box size in the
electron drift direction $y$ to achieve convergence. A box resolution
down to $\rho_s$ guarantees inclusion of all scales necessary for
the nonlinear dynamics that are esssential for the drift wave turbulence
characteristics.

\begin{figure}[H]
\includegraphics[width=8.5cm]{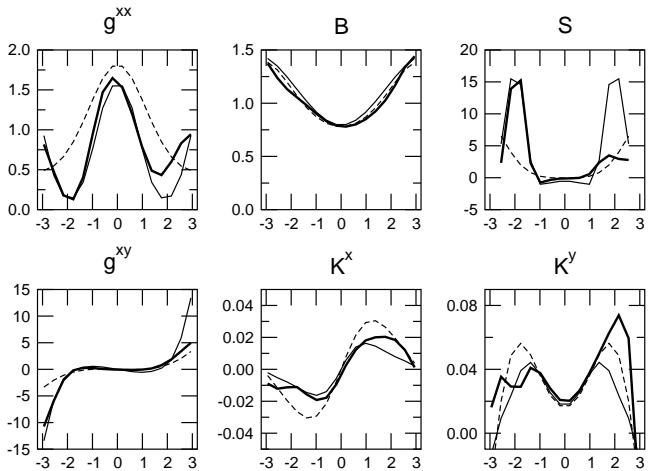}
\caption{\label{f:gxx} Metric element $g^{xx}$, $g^{xy}$ (unshifted),
normal curvature ${\cal K}^y$, geodesic curvature ${\cal K}^x$,
magnetic field strength $B$ and local magnetic shear $S$ in a tokamak
with elongation $\kappa=1$ and triangularity $\delta=0$ (dashed lines),
compared to a configuration with $\kappa=2$ and $\delta=0.4$ (thin
solid lines) and to an actual ASDEX Upgrade configuration (bold solid
lines).}  
\end{figure}

\section{Flux tube representation of tokamak geometry}

Tokamak equilibria are computed by solving the Grad-Shafranov /
L\"ust-Schl\"uter equation with the code HELENA \cite{HELENA},
implemented as described in Ref.\ \cite{Scott2000}.  A set of
nested flux surfaces in straight field line Hamada coordinates $(V,
\theta, \zeta)$ is obtained by specification of given radial profiles
of pressure and rotational transform, and of the shape of the bounding
last closed flux surface. These Hamada coordinates are then
transformed into a field aligned system and re-scaled into local flux
tube coordinates $(x,y,z)$ \cite{ScottVar98}.  Global consistency in the
parallel boundary condition for one connection length is maintained
\cite{Scott98}.
A transformation of the coordinate $y$, which signifies the electron
diamagnetic drift direction perpendicular to the magnetic field within
flux surfaces, is applied in order to avoid grid deformation by local
magnetic shear \cite{Scott01}. Otherwise, grid cells are sheared
strongly in $y$ direction particularly near the X-point region, with
malign consequences on nonlinear dynamics, especially in the vorticity,
that lead to a violation of the basic drift wave character and
overestimate linear MHD-like dynamics.
The differential operators are then expressed in terms of the flux
tube coordinates: The curvature operator becomes
\begin{equation}
{\cal K} = {\cal K}^x(z) \nabla_x + {\cal K}^y(z) \nabla_y,
\end{equation}
the perpendicular Laplacian in flute mode ordering is
\begin{equation}
\nabla_{\perp}^2  = g^{xx}(z) \partial_{xx} + 2 g^{xy}(z)
\partial_{xy} + g^{yy}(z) \partial_{yy},
\end{equation}
and the parallel derivative is
\begin{equation}
\nabla_\parallel   = b^z(z) {\partial_z},
\end{equation}
noting that the factor of $B^2$ in $\rho_i^2$ also depends on $z$.
Some metric coefficients $g^{ij}$ that were obtained for elongation
$\kappa=1$ and $\kappa=2$ with triangularity $\delta=0$ and
$\delta=0.4$ are shown in Fig.~\ref{f:gxx}. Increasing elongation
$\kappa$ specifically rises the local magnetic shear $S = \nabla_{\parallel}
( g^{xy}/g^{xx} )$ and reduces ${\cal K}^x$  both in the upper and
lower regions of the torus that correspond to flux tube coordinates
$z=\pm \pi/2$. In the following, the configuration with $\kappa=1$ and
$\delta=0$ will be refered to as a ``simple circular torus'' (SCT).

Local and global magnetic shear are in general known to have a damping
influence on tokamak edge turbulence \cite{Kendl03}, whereas geodesic
curvature acting through ${\cal K}^x$ upon the axisymmetric component
($k_y=0$ components, or ``modes'') maintains
the coupling for a loss channel from zonal flow energy eventually to
turbulent
vortices \cite{Scott03}. Both mechanisms help to reduce turbulent
transport. Normal curvature ${\cal K}^y$ on the other hand strengthens
primarily the interchange forcing of the turbulence ($k_y \neq 0$).

\section{Transport scaling by elongation and triangularity}

A series of tokamak equilibria is constructed for elongation $1 \leq
\kappa \leq 2$ and triangularity $0\leq \delta \leq 0.4$
with equal profiles of pressure and rotational transform.
The flux tube is chosen at a radial position $r = \sqrt{V/V_0}=0.95$,
where $V_0$ is the volume enclosed by the last closed flux surface.
The background density profile and rotational transform for the GEM turbulence
computations are linearised within the bounds of the radial computational
domain. We restrict our simulations for now to the closed field line region
lying a few tens of ion gyroradii inside from the separatrix, thus avoiding
complications that occur by a divergent metric and associated grid
deformation when the last closed flux surface is approached in the vicinity of
an X-point.
Ultimately, the goal of edge turbulence simulations will be to combine
sufficiently well resolved nonlocal drift wave computations with a
representation of the realistic field line geometry crossing the
separatrix to the bounded scrape-off layer region.

When the plasma shape is thus varied, we have a particular interest in
the effects on fluctuation time and spatial scales, on the radial
variation of flux surface averaged (zonal) flows, and on turbulent
transport.
We first apply the ensemble of shaped tokamak equilibria to the DALF model.
The cross-field turbulent electron transport is determined by the flux
surface average of the radial $E \times B$ convection of
the fluctuating density by $F_e = \langle \tilde v_x \tilde n_e
\rangle_{yz}$, given in standard gyro-Bohm normalisation
to $n_e c_s (\rho_s / L_{\perp})^2$.
The transport flux $F_e(t)$ is fluctuating in time, and for
specifying a quantitative value to it the time average over a
sufficiently long window, that covers all relevant frequency scales,
is taken during the final phase of simulations after all initial
linear transients and spin-up of flows and oscillations have reached
saturation.

\begin{figure}[H]
\includegraphics[width=7.5cm,height=5.5cm]{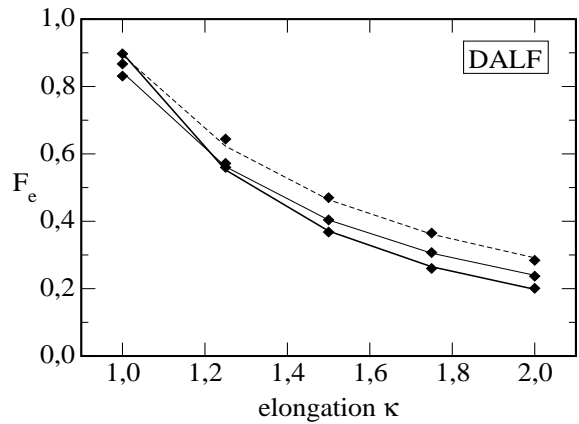}
\caption{\label{f:kap-dalf} DALF: Dependence of turbulent electron transport
$F_e$ (in gyro-Bohm units) on elongation $\kappa$ for triangularity
  $\delta=0$, $0.2$ and $0.4$ (from bottom to top).}  
\end{figure}

\begin{figure}[H]
\includegraphics[width=7.5cm,height=5.5cm]{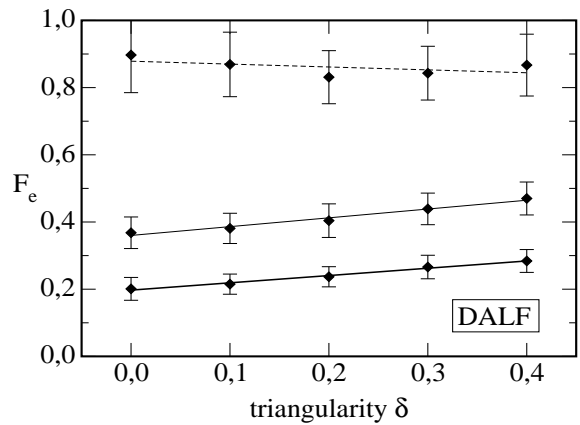}
\caption{\label{f:del-dalf} DALF: Dependence of turbulent electron transport
$F_e$ (in gyro-Bohm units) on triangularity $\delta$ for elongation
  $\kappa=1.0$, $1.5$ and $2.0$ (from top to bottom).}
\end{figure}

As shown in Fig.~\ref{f:kap-dalf}, we find that $F_e$  is reduced to
around a third when elongation is increased
from a circular cross section to $\kappa=2$. Scaling $F_e$ against
triangularity $\delta$, we find in Fig.~\ref{f:del-dalf} that for $\kappa=1$
the transport flux is independent from $\delta$ within the error
bars given by the deviation of the fluctuating data from its time average.
For higher elongation $\kappa=2$ the transport $F_e$
is increased from the normalised value of $0.20$ to $0.28$ by
increasing $\delta$ from $0$ to $0.4$.
A good fit to the transport data in the DALF model within fluctuation error
bars is obtained by
\begin{equation}
F_e^{D\!A\!L\!F} (\kappa, \delta) \approx F_{1,0} \cdot \kappa^{-2.15 +
1.46 \delta}
\label{eqkappa}
\end{equation}
where $F_{1,0} = F (\kappa\!=\!1, \delta\!=\!0)$.
This reduction of transport by elongation can be related to three different
causes: 

(1) Increase of local and global magnetic shear (magnetic shear
damping), (2) reduction of curvature drive (interchange coupling), and
(3) an enhancement of sheared zonal flows (geodesic transfer).

In Section VI we will present results showing that the effect of flow
shear enhancement
by a weakening of the geodesic transfer mechanism is insignificant for DALF
parameters. Also for these resistive electromagnetic edge parameters, the
interchange dynamics in the normal curvature forcing is acting more as a
catalyst than as a drive mechanism for the turbulence, with only weak
ballooning of the fluctuating quantities observed. On the other hand, it has
been established that local magnetic shear by flux surface deformation has a
strong damping influence on tokamak edge turbulence: comparing the case of a
simple circular torus (with global shear only) to a case where the local
magetic shear structure is typical of a shaped tokamak (but the curvature
properties are those of a circular torus), a reduction of transport to
around
a half was found \cite{Kendl03}.
Moreover, the effective topological global shear (defined, e.g., in
Eq.~29 of Ref.~\cite{Scott98}) is also increased by finite aspect ratio
and elongation, but is not significantly affected by moderate
triangularity ($\delta < 0.6$ or so).
It can thus be argued, that the major
mechanism for reduction of transport by elongation in the DALF case is
magnetic shear damping, with an additional (slightly lesser) influence by
interchange drive reduction, and an only insignificant zonal flow
enhancement.

We now include ion temperature gradient (ITG) driven dynamics by
allowing for
a gradient ratio $\eta_i=2$ and using the full six-moment version of GEM.
In the ITG model, the flux-surface averaged turbulent transport can be
characterised by particle transport $F_n = \langle \tilde n \tilde v_x
\rangle$ and by heat transport  
$Q_i = Q_i^{cv} + Q_i^{cd}$ with convective component $Q_i^{cv} = {3 \over
  2} \tau_i \langle \tilde n_i \tilde u_x + \tilde T_{i\perp} \tilde w_x
  \rangle$
and conductive component $ Q_i^{cd} = \tau_i \langle ({1 \over 2} \tilde
T_{i||} + \tilde T_{i\perp}) \tilde u_x + (\tilde n_i + 2 \tilde T_{i\perp})
\tilde w_x \rangle$, where $\tilde u_x$ is the gyro-averaged ExB
velocity and
$\tilde w_x$ is its FLR correction.
Normalisation is to standard gyro-Bohm units with the gradient scale length
set to unity, and $\tau_i=T_i/T_e$ is the temperature ratio.

\begin{figure}[H]
\includegraphics[width=7.5cm,height=5.5cm]{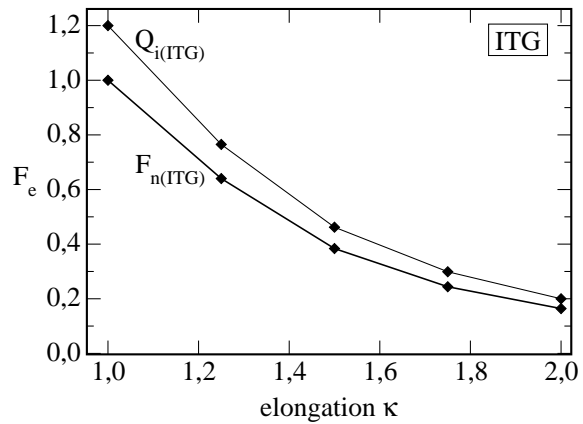}
\caption{\label{f:kap-itg} ITG: Dependence of turbulent particle transport
$F_n$ and heat transport $Q_i$ (in gyro-Bohm units) on elongation
$\kappa$ for
  triangularity $\delta=0$.}
\end{figure}

\begin{figure}[H]
\includegraphics[width=7.5cm,height=5.5cm]{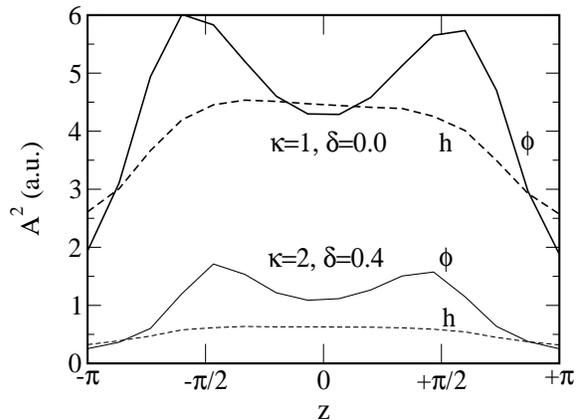}
\caption{\label{f:balloon} ITG: In the shaped configuration with $\kappa=2$
  the coupling quantity $h_e(z) = \tilde p_e(z) - \tilde \phi(z)$ has a
  reduced squared amplitude compared to $\tilde \phi(z)$, indicating a
weaker
  interchange ballooning character of the turbulence than in a circular
torus.}
\end{figure}

When displaying heat and particle transport in relation to elongation and
triangularity now for the ITG model, we find in Fig.~\ref{f:kap-itg} a
scaling
of $Q_i \sim \kappa^{-2.6}$ and $F_n \sim\kappa^{-2.3}$. The reduction by
elongation is thus stronger than for the DALF model. One evident
mechanism for
this increased reduction in the ITG case is the weakening of the interchange
drive of the ITG instability acting via the normal curvature ${\cal K}^y$,
which is slightly reduced by elongation in the ballooning region of the flux
tube (compare Fig.~\ref{f:gxx}). In contrast to DALF dyanmics, the ITG
turbulence is strongly driven by ``unfavourable'' normal curvature,
leading to a pronounced ballooning character of the fluctuations, as
shown in Fig.~\ref{f:balloon}.

We find as an additional mechanism in the ITG case a significant weakening of
the geodesic transfer mechanism mediated by ${\cal K}^x$, which is discussed
in the following section.

\begin{figure}[H]
\includegraphics[width=7.5cm,height=5.5cm]{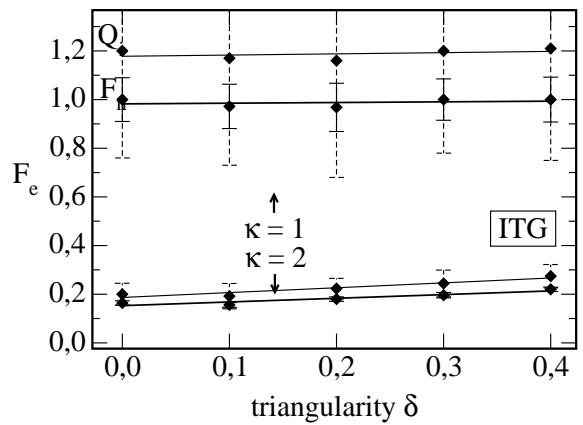}
\caption{\label{f:del-itg} ITG: Dependence of turbulent particle transport
$F_n$ and heat transport $Q_i$ (in gyro-Bohm units) on triangularity
$\delta$ for elongation $\kappa=1$ and $2$.}
\end{figure}

Similar as in the DALF case, both heat and particle transport are slightly
increasing with higher triangularity for an elliptical cross section
($\kappa=2$) in Fig.~\ref{f:del-itg}, while for small $\kappa$ no influence
from triangularity is observed within the fluctuation error bars.

Summarising the results of this section on the influence of flux surface
shaping on turbulent transport for different edge parameters, we find
that the main contribution to transport reduction in both the DALF and ITG
regimes is magnetic shear damping by elongation.  

Shear flow enhancement by the reduced geodesic transfer mechanisms is found to
be relatively weak in the DALF case, but is becoming relevant in the ITG
model, as can bee seen from analysing the wave number and frequency spectra
discussed in the next section.

For fixed gradients, the local diffusivity is proportional to the transport
and thus to the roughly inverse quadratic scaling with $\kappa$ given
above. Global transport across an entire magnetic flux surface can in
principle be obtained by integration of the local transport
coefficients, that takes the variation of surface area with elongation into
account. 

\section{Frequency and wavenumber spectra of flows and fields}

Time traces are resolved down to $1/20$th of the drift
frequency $\omega_{DW} = c_s / L_{\perp}$, and long computational
runs are required to account for zonal flows with $\omega \approx
0$. A typical number of time steps is $10^5$. 

Geodesic acoustic oscillations (GAM) may be distinctly detectable for some
parameters and are expected for circular geometry around $\omega / \omega_{DW}
= (2 L_{\perp} / R) \sqrt{(1+ \tau_i)/2}$. In the DALF case for $\tau_i=0$ we
have $\omega_{GAM} = 0.035 \omega_{DW}$, and in the ITG case for $\tau_i=1$
we have $\omega_{GAM} = 0.05 \omega_{DW}$. 

In general, the GAM frequency is determined by the geodesic curvature term
${\cal K}^x$ and is influenced by flux surface deformation.  
The geodesic curvature effect is the action of the curvature operator
${\cal K}$ upon the axisymmetric component ($k_y=0$).
In local flux tube coordinates we find ${\cal K}^x \sim (1 / \kappa)$, and
the factor $1/\kappa$ in the geodesic curvature term does also
accordingly scale the geodesic acoustic frequency: stronger
elongation shifts the GAM resonance frequency in the spectrum closer
to the zero-frequency zonal flow. The same argument, now applied to
the normal curvature term ${\cal K}^y$, also accounts for a
$\kappa^{-1 }$ scaling of the (interchange) drift wave frequency.

\begin{figure}[H]
\includegraphics[width=7.5cm,height=5.4cm]{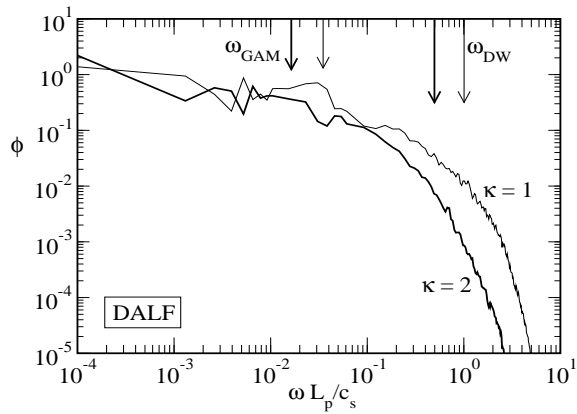}
\caption{\label{f:freq-dalf} DALF: Local frequency spectra $\tilde \phi
  (\omega)$
  for $\kappa=1$ (thin line) and $\kappa=2$ (bold line), both for
  $\delta=0$. Arrows indicate the drift wave and GAM resonances for
  $\kappa=1$.}  
\end{figure}

For DALF parameters used in our computations, the GAM peak does not
protrude very distinctly but can still be identified on the flat top
of the spectra.  In Fig.~\ref{f:freq-dalf} the discrete Fourier transform
spectrum $\tilde \phi(\omega)$ for a
measurement of $\tilde \phi$ at a local point in the centre of the
$xyz$-domain is shown: the spectrum below the GAM
resonance is mostly flat except for the distinct zonal flow peak at
$\omega=0$. The higher frequency part
shows a typical $\omega^{-\alpha}$ cascade structure, where the
exponent $\alpha$ changes around the drift frequency
$\hat \omega \approx 1$, which for our choice of edge parameters
coincides also with the Alfv\'en frequency. 

One may identify three ranges in the spectrum: A flat top region between the
zonal flows and GAM frequencies, one intermediate range between the
GAM and drift wave frequencies with $\alpha \approx 1-2$, and a
cascade range between the drift wave and the dissipation times scale
with $\alpha \approx 3-5$.
As expected, both GAM and drift wave frequencies are reduced by
increasing elongation to lower values $\omega \rightarrow \omega /
\kappa$. The whole spectrum is thus shifted to the left for $\kappa=2$
in Fig.~\ref{f:freq-dalf}.  

\begin{figure}[H]
\includegraphics[width=7.5cm,height=5.4cm]{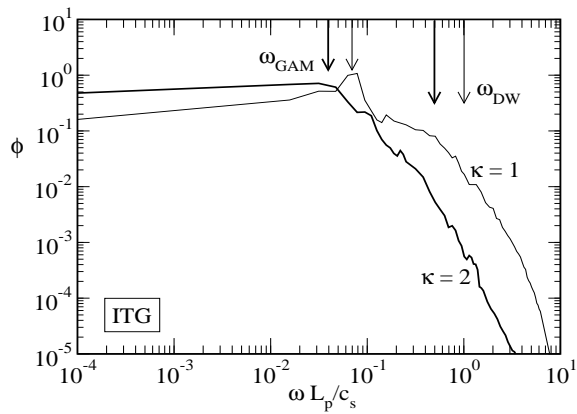}
\caption{\label{f:freq-itg} ITG: Local frequency spectra $\tilde \phi
(\omega)$
  for $\kappa=1$ (thin line) and $\kappa=2$ (bold line), both for
  $\delta=0$. Arrows indicate the drift wave and GAM resonances for
  $\kappa=1$.}  
\end{figure}

Using ITG parameters, the GAM peak is more clearly visible for the circular
torus case but vanishes for elongation $\kappa=2$, as can be seen in
Fig.~\ref{f:freq-itg}. The whole spectrum experiences again a shift
to lower frequencies for the elongated case. The $\omega=0$ zonal flow
component in the ITG spectrum exhibits significantly higher amplitude for
elongation $\kappa=2$ than for the circular torus.

In addition to computing the flux surface shapes with varying elongation
and triangularity discussed above, HELENA is also used to obtain
equilibria with an outer boundary defined by an experimentally
reconstructed separatrix position. In this way we obtain an
equilibrium in the typical shape of an ASDEX Upgrade (AUG) \cite{AUG}
lower single null divertor configuration (as shown in Fig.~\ref{f:aug}).
The additional asymmetric shaping by the presence of an X-point in the
lower part of the torus leads to an increase of local magnetic shear
near $z=-\pi/2$ and an associated local reduction of geodesic curvature.
The elongation and triangularity of this configuration are otherwise
comparable to the kappa-delta model with
$\kappa \approx 1.6$ and $\delta=0.3$. This AUG
configuration was first used in edge turbulence computations
in Ref.~\cite{ScottVar98}, where the dependence of transport on $\hat
\beta$ was established:
it was found that the onset of MHD ballooning mode turbulence is
prevented for typical tokamak edge parameters due to the shaping
effects in realistic geometry, and the nature of transport is still
basically of the drift-Alfv\'en wave character.

The AUG model has, for the present flux tube position,
properties relevant to the turbulence dynamics that are in some
aspects in between those of up-down symmetric configurations with
$\kappa=1.6$ and $2.0$. Transport in the AUG model is for DALF parameters
found to be reduced to $F_e = 0.5$ compared to $F_e=0.9$ for the circular
torus. The nonlinear DALF growth rate $\Gamma_N = F_e / 2 E_{{\rm tot}}$ is
only slightly lower for the AUG model with $\Gamma_N=0.007\pm0.001$ than
for the circular torus with $\Gamma_N=0.0085\pm0.001$.

For ITG parameters, the particle transport is nearly identical in the
AUG and
SCT cases, both at $F_e = 1.0$. The nonlinear ITG growth rate is also
similar, $\Gamma_N=0.0080\pm0.0020$ for AUG and
$\Gamma_N=0.0088\pm0.0024$ for
the circular torus.

A reduction of geodesic curvature by elongation and X-point shaping
might be expected to lead to a weakening of the geodesic transfer
coupling mechanism for energy from zonal flows to GAMs \cite{Kendl05}.
This can be analysed by comparing the flux surface averaged vorticity
$\langle \Omega_E (x) \rangle = \partial_x \langle v_E^y (x) \rangle$
for the configurations of the circular torus with the AUG model.
The zonal vorticity $\langle \Omega \rangle / \Gamma_N$, set in relation
to the
nonlinear drive rate $\Gamma_N$, thus represents the radial shearing of
zonal
flows, which is considered responsible for the turbulent shear flow
decorrelation and energetic damping of vortices \cite{Diamond94}.

Concerning the geometric effect on this, we find different results for the
DALF and ITG parameters.

\begin{figure}[H]
\includegraphics[width=7.5cm,height=5.5cm]{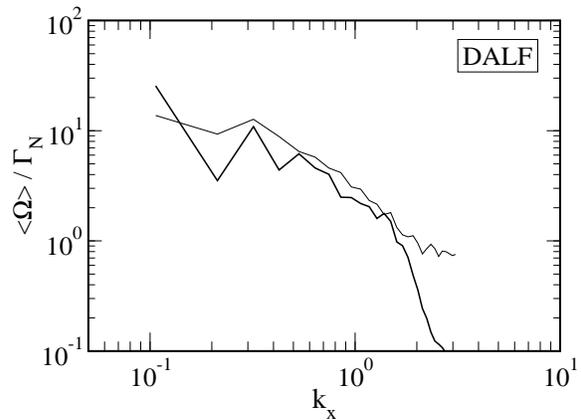}
\caption{\label{f:zf-dalf} DALF: $k_x$ spectra of flow shear rate
  in AUG geometry (bold line) and for the simple circular torus (thin
  line). The cumulative flow shear amplitude in the spectra is similar for
  both cases.}
\end{figure}

In the DALF case, the cumulative amplitude of the components in the radial
spectrum
of flow shear $\langle \Omega \rangle / \Gamma_N$ is the same within
fluctuation error bars for AUG geometry as for the circular torus.
The lowest $k_x$ component in these spectra, which is found to be
stronger in the AUG case, makes up for only approximately 8\% of the total
vorticity. The spectra are shown in Fig.~\ref{f:zf-dalf}.
Flow shear in the DALF case thus is not significantly enhanced by flux
surface shaping. Also the amplitude of zonal flows in the DALF frequency
spectra in Fig.~\ref{f:freq-dalf} is found to be comparable for circular and
shaped configurations. The reduction of turbulent edge transport is in this
case therefore mainly a result of the local magnetic shear effect
\cite{Kendl03}.  

\begin{figure}[H]
\includegraphics[width=7.5cm,height=5.5cm]{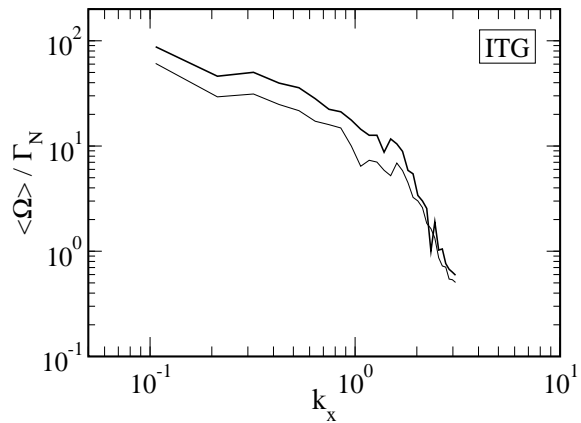}
\caption{\label{f:zf-itg} ITG: $k_x$ spectra of zonal vorticity
  related to nonlinear growth rate (setting the flow shear) in AUG and SCT
  geometry. Shear flows are cleary enhanced in the AUG configuration in all
  components.}  
\end{figure}

In the ITG case, the flow shear rate deduced from the $k_x$ spectra of zonal
vorticity is cleary higher in all components for the AUG tokamak than
for the
circular torus, as can be seen in Fig.~\ref{f:zf-itg}.
The cumulative amplitude in the spectra is by 70\% higher in AUG geometry
than in simple torus geometry. Flux surface shaping introduced by elongation
and X-point shaping clearly enhances flow shear significantly for ITG edge
parameters. The zonal flow amplitude in Fig.~\ref{f:freq-itg} is also by a
factor of three larger in the elongated configuration.

This shear flow enhancement contributes to the overall turbulent transport
reduction by elongation, as has been discussed in section V, in addition to
the effects of magnetic shear damping and the reduction of interchange
drive. The detailed relative importance of these three mechanisms may
however change with varying plasma parameters.

\section{Conclusions and outlook}

We have discussed the effects of flux surface shaping on tokamak edge
turbulence and flows. Elongation strongly reduces the turbulent transport,
whereas triangularity was found to have only weak (transport enhancing)
influence. Two regimes of edge plasma parameters were studied: an
electromagnetic resistive drift wave (DALF) case, and the (ITG) case with an
additional ion temperature gradient of $\eta_i=2$.

The scaling of local transport is roughly with the inverse square
of elongation, showing a stronger reduction for ITG parameters than for DALF
parameters.  The ITG result is consistent with earlier findings with the
previous version of the GEM model \cite{Scott2000}.

The mechanisms of flux surface shaping effects on turbulence have been
discussed. The major cause for transport reduction by elongation is
damping of turbulence by stronger local magnetic shear at the upper
and lower parts of the torus on top of increased global magnetic shear.
The supervening reduction of the interchange drive by a lower normal
curvature has an additional turbulence reduction effect in particular
for ITG
parameters, where the interchange drive is slightly weakend.
In the ITG case a significant enhancement of zonal flow shear by
elongation and X-point shaping further contributes to transport reduction.
The shear flow effect is on the other hand found to be negligible for DALF
parameters.

Both the DALF and ITG parameter regimes used in our computations seem
consistent with the presently still poor experimental knowledge on edge ion
temperature profiles, though this situation is rapidly improving
\cite{Reich04}. It can be assumed that additional core plasma
heating, e.g. by neutral beam injection (NBI) used in particular for
obtaining
the transition to an H-mode, is both increasing the ion temperature at the
pedestal top and the ion temperature gradient in the overall edge pedestal
region. Based on our results, the shaping of tokamak flux surfaces by
elongation and by the additional presence of an X-point like in AUG is
expected to facilitate an enhancement of zonal flows when the role of
the ion
temperature gradient dynamics is pronounced by additional heating.
  
It can be expected that this enhancement of zonal flows found for ITG
parameters is even more pronounced for more strongly shaped flux surfaces
nearer to the separatrix. Preliminary results suggest that such initial
radially local flows have the ability to spread over wider regions of the
pedestal in simulations with global profile evolution.

Some L-H transition theories include zonal flows as a trigger mechanism to
induce the transition to a sustained mean shear flow. The enhancement of
zonal
flows by flux surface and X-point shaping in the presence of ion temperature
gradient steepening may offer an additional explanation for the observation
that H-mode access in tokamak experiments is facilitated by the presence
of a
divertor.

The possibility to obtain a confinement transition within first principle
computations of edge turbulence will thus have to be studied with a code
that
includes full temperature dynamics, realistic flux surface geometry, global
profile evolution, and a coupling of edge and SOL regions including
realistic
sheath boundary conditions, while in addition it maintains sufficient
grid resolution, grid deformation mitigation, and energy plus enstrophy
conservation in the vortex/flow system.

\section*{Acknowledgement}

This work was supported by the European Commission under
contract FU06-CT-2003-00332.

\par\vfill\eject

\end{document}